# Critical role of parallel momentum in quantum well state couplings in multi-stacked nanofilms: an angle resolved photoemission study

Woojoo Lee[1], Chi-Ruei Pan[2], Hyoungdo Nam[1], Mei-Yin Chou[2,3,4] and Chih-Kang Shih[1,*]

[1]Department of Physics, The University of Texas at Austin, Austin, Texas 78712, USA.

[2] School of Physics, Georgia Institute of Technology, Atlanta, Georgia 30332, USA

[3] Institute of Atomic and Molecular Sciences, Academia Sinica, Taipei 10617, Taiwan

[4]Department of Physics, National Taiwan University, Taipei 10617, Taiwan

*Corresponding author: shih@physics.utexas.edu

## Abstract

We use angle resolved photoemission spectroscopy (ARPES) to investigate the coupling of electron quantum well states (QWS) in epitaxial thin Pb and Ag films. More specifically, we investigate the Ag/Si, Pb/Si, and Pb/Ag/Si systems. We found that the parallel momentum plays a very profound role determining how two adjacent quantum wells are coupled electronically across the interface. We revealed that in the Pb/Ag bimetallic system, there exist two distinctly different regimes in the energy versus momentum ($E\ vs.\ k$) space. In one regime the electronic states in Ag and Pb are strongly coupled resulting in a new set of quantum well states for the bi-metallic system. In the other regime the electronic states in individual metallic layers are retained in their respective regions, as if they are totally decoupled. This result is corroborated by calculations using density functional theory. We further unravel the underlying mechanism associated with the electron refraction and total internal reflection across the interface.

**Introduction**

Quantum well systems have played an important role in contemporary condensed matter physics and have been a key enabler for modern electronics. In a quantum well system, its electronic states can be designed through the control of the boundary conditions[1–3]. Coupling different quantum well systems together further add to the design complexity[4–8]. This approach has enabled researchers to create "designer electronic structures" providing a rich quantum material platform for realizations of novel physical phenomena and new technology[9–15].

In designing coupled quantum well systems, the most critical factor is to understand how the quantum well states (QWS) in different layers are coupled to each other and how this coupling leads to the resulting electronic structures of the composite as a whole. For example, some aspects of bi-quantum well systems (e.g. Pb/Ag and Ag/Si) have been studied previously. In the case of Ag/Si, several earlier studies reported observation of the coupling between the Schottky barrier (SB) confined Si hole-like QWS and the QWS of thin Ag films[7,16–20]. In the case of Pb/Ag system, it was discovered that the electrons in the buried Ag layer can transmit coherently across the interface through the Pb overlayers and be detected as photoelectrons (even though the overlayer thickness is larger than the electron escape depth)[8]. Nevertheless, the role of parallel momentum in a coupling of these systems has rarely been discussed in previous ARPES (angle-resolved photoemission) experiments, despite its importance in governing the electrical properties of materials.

Here we use angle-resolved photoemission spectroscopy (ARPES) to investigate the behavior of asymmetric double quantum well systems Pb/Ag. Several new aspects have been uncovered in the current study. More specifically, we unveil the critical role of the parallel momentum, $k_\parallel$, in determining the coupling of adjacent quantum well systems. Especially, we show that near the Brillouin zone (BZ) center the Ag QWS are well coupled to the Pb QWS, forming a composite Pb/Ag QW system. On the contrary, states in the BZ regions with large parallel momentum are completely decoupled from the Ag layer, retaining the original E vs k dispersion of the Pb quantum well system. Similarly, for Ag QWS with a large crystal momentum, they are retained in the original Ag layer and could not transmit through the Pb overlayer. We attribute these behaviors to the electron refraction across a potential step at the Pb-Ag interface. These findings are further corroborated using first-principles calculations within density functional theory (DFT).

**Experimental details**

ARPES measurement was performed with the sample held at liquid nitrogen temperature. Helium lamp light (21.2eV and 40.8eV) was used as a photon source and the spectra was acquired using a Scienta R3000 analyzer. The pressure was maintained under $2 \times 10^{-9}$ torr during Helium lamp operation.

The samples were grown in MBE chamber under $2 \times 10^{-10}$ torr pressure by evaporating Ag, Pb source from K -cell and transferred to ARPES chamber via our Ultra High Vacuum(UHV) transfer vessel (based pressure < $1\times 10^{-9}$ torr). All metal overlayers are grown on an atomically clean Si(111) 7x7 substrate (prepared by typical Si-flashing procedure). For epitaxial Ag/Si and Pb/Si growth, 2 step method is used. In the case of Ag/Si, clean Si(111) 7x7 substrate is cooled down to 100K and Ag source is evaporated on it with 0.3-0.5ML/min deposition rate. And next, it is annealed at room temperature for one hour. Lastly, it is cooled down to 100K for the measurement. In the case of Pb/Si, a stripped incommensurate (SIC) wetting layer is first prepared on a clean Si(111) 7x7 substrate by depositing over 1.3ML Pb at room temperature and post-annealing it at 400-450°C for 4 mins. And next, the sample is held at 100K and Pb is deposited on it with 0.3-0.5ML/min deposition rate. Lastly, the sample is annealed at 250K for 15min and then cooled down to 100K immediately to prevent the de-wetting process of Pb layers. In the case of Pb/Ag/Si, we used the ARPES measured Ag/Si sample made by the above procedure. The sample is held at liquid nitrogen temperature and Pb is deposited on the Ag/Si with 0.3-0.5ML/min deposition rate, the sample is post-annealed at 250K for 3 hours. Lastly, it is cooled down to 100K for measurements [13,15,21–23].

The crystal orientations are determined using low energy electron diffraction. In all cases, the epitaxial metal films (Ag, Pb or Pb/Ag) are all rotationally aligned with the Si(111) substrate. The lattice constant ratio for Pb:Si (Ag:Si) is 10:11 (3:4). The ratio between Pb:Ag is 40:33, which is close to 6:5 (to within 1%).

## Computational details

Density functional theory (DFT) calculations were performed with the Vienna ab-initio Simulation Package (VASP)[24,25] using the plane wave basis and an energy cutoff of 500 eV. The projector augmented wave (PAW)[26] method and the exchange-correlation functional of the Perdew-Burke-Ernzerhof (PBE) form[27] in the generalized gradient approximation (GGA) are employed. The supercell lattice constant of the Pb/Ag bimetallic structure was chosen to be an average supercell lattice constant based upon 6× 6 Pb(111) and 5×5 Ag(111) (i.e., $L_{supercell} = (5 \times a_{Pb(111)} + 6 \times a_{Ag(111)})/2$), so that the strain levels are -0.3% and 0.3% for Pb and Ag, respectively. The self-consistent cycles were carried out until the energy difference between two cycles was less than $10^{-4}$ eV. The supercell electronic structure was unfolded in reference to the Pb Brillouin zone using the BandUP code[28–30].

## Results and discussion

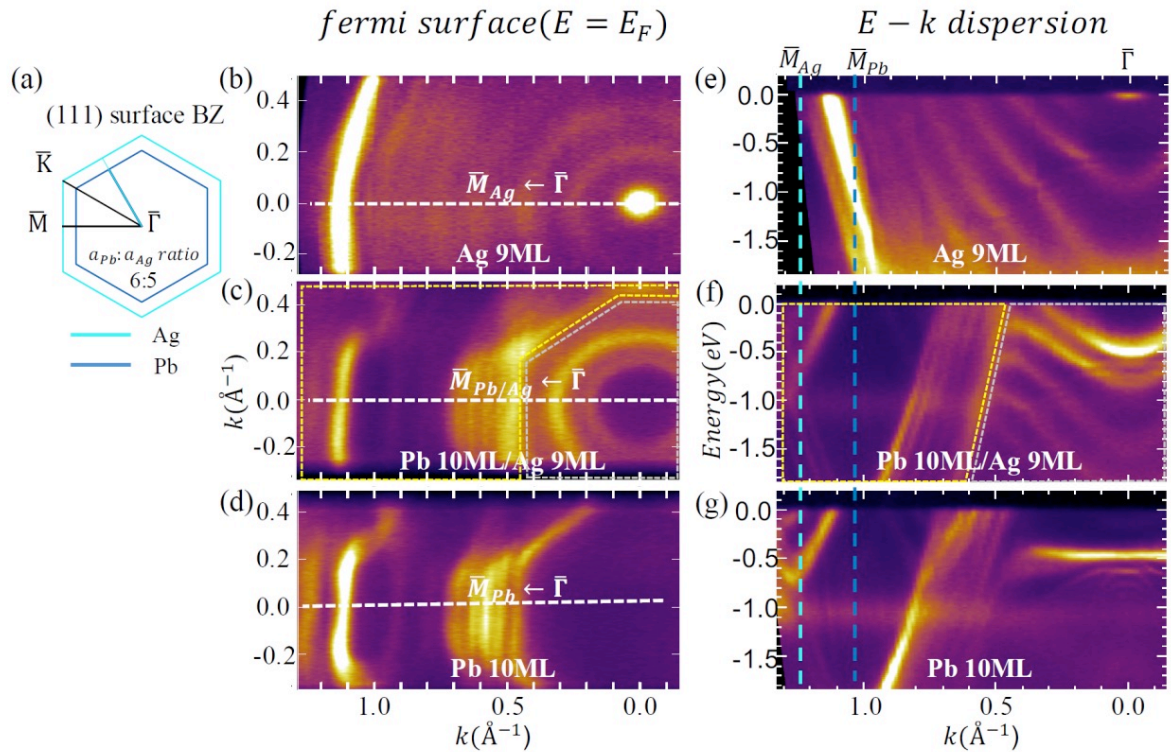

*Figure 1. ARPES measurement on Ag 9ML, Pb 10ML/ Ag 9ML and Pb 10ML at liquid nitrogen temperature with $\hbar\omega = 21.2 eV$. (a) Surface Brillouin zone of Pb (111) and Ag (111) films. ARPES mapping of the fermi surfaces (b, c, d) and the band structures (e, f, g) for Ag 9ML, Pb 10ML/Ag 9ML and Pb 10ML. The data is taken along the $\bar{\Gamma} - \bar{M}$ direction. White dashed lines in (b, c, d) indicate the $\bar{\Gamma} - \bar{M}$ direction. Light and dark blue dashed lines in (e, f, g) indicate M points of Ag and Pb respectively. Two distinct*

.

The fermi surface and E-k dispersion mapping images of Ag 9ML, Pb 10ML/Ag 9ML and Pb 10ML are represented in Fig. 1 taken from photoemission measurements. We first discuss results for individual Ag or Pb layers on Si(111). In Ag/Si, the QWS, well-resolved in spectra are obtained. The parabolic shape bands in E-k dispersion indicate quasiparticles in Ag films behave like free electrons. In addition, one also observes fringes due to the coupling of Ag QWS to the Si downward hole QWS confined by the Schottky barrier at the Ag/Si interface. These features are well-understood and have been reported several times previously[7,16–20].

For Pb/Si, the spectra show rather flat QWS at $E_B$~ -0.5 eV near the zone center with an effective mass of significantly larger than the result using DFT calculation. As previously reported, the strong mass enhancement can be attributed to two different factors; the coupling to the VBM of the underlying Si and electron localization induced by lateral atomic spacing of Pb layers[31,32]. Other QWS near the zone center can be better observed using He-II excitation as shown in Fig. S1 accompanied by the energy distribution curve (EDC) at the Γ point. The coupling of this QWS (-0.5 eV) with the SB confined hole QWS of Si can also be observed as fringes in Fig. 1g.

In addition to QWS near the zone center, we observe several well-separated bands with negative linear dispersion in the approximated region with k = 0.5 – 0.9 Å$^{-1}$. These states correspond to the folded-back QWS (see also DFT calculation shown in Fig. 3a). The DFT calculation shows many closely spaced folded back quantum well states and many of them are nearly degenerate. Experimentally, we can observe up to 6 well separated bands.

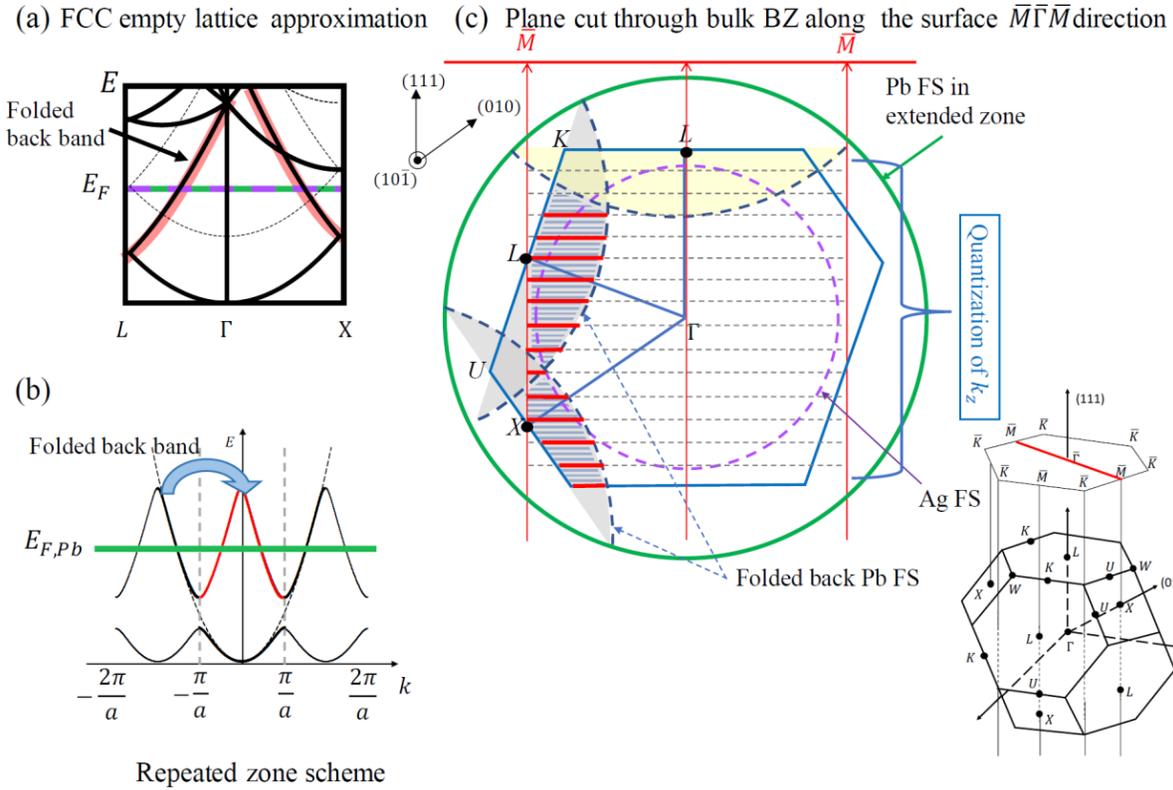

*Figure 2. Origin of the folded back bands and Brillouin zone analysis. (a) fcc empty lattice approximation for Pb(solid line) and for Ag(dashed lines). Red shaded lines are corresponding to folded back bands. Which shape can be understood with 1D nearly free electron model in (b). Note that different bands in (a) inherently come from $E = \frac{\hbar^2}{2m_0}(\vec{k} + \vec{G})^2$, $\vec{G}$ is a reciprocal lattice vector of fcc lattice. (c) Side view of fcc Brillouin zone along $(10\bar{1})$ direction. Pb and Ag Fermi surfaces are represented by a green solid line and a purple dashed line respectively. Grey dashed lines are drawn to indicate quantized $k_z$ momentum caused by the quantum confinement effect. Red solid lines are overlaid on the grey color folded back Pb FS area in order to explicitly show that the Pb's negative linear dispersion bands are originated from folded free electron like bands. The folded back area along the (111) direction is shaded by yellow color. The bands in this area are corresponding to parabolic shape bands near the Fermi energy in the Pb band structure.*

The existence of the folded back QW bands in Pb band structure can be understood with nearly free electron model within the fcc (face-centered-cubic) empty lattice approximation. In Fig. 2a, empty lattice approximation shows E-k dispersion along the $\bar{\Gamma} - \bar{L}$ and $\bar{\Gamma} - \bar{X}$ direction in fcc structure; E-k dispersion along each direction is associated with 1D nearly free electron model with different lattice constant. Notably, it includes negative linear dispersion bands. As can be seen in Fig. 2b, these bands are corresponding to folded back bands of nearly free electron model. In the case of Pb, since it contains 4 free electrons, it has a larger fermi sphere than Ag which has only 1 free electron. In that Pb's fermi level can be located in the region where the folded back bands exist, on the contrary, Ag's fermi level only can fill the parabolic shape bands. Since our ARPES result is based on surface projected BZ, to clarify how the quantized bands are formed, we draw a bulk BZ diagram on Fig. 2c. Due to the larger fermi surface of Pb, its green circle is folded back into the BZ rather than Ag's purple circles does not. Along the (111) direction, there are many $k_z$ quantized states in the BZ. Red overlaid solid lines in the grey shaded area are associated with negative linear dispersion folded back bands and grey dashed in yellow shaded area are associated with parabolic shape bands near the Γ point (Fig. 3a).

The most interesting features are in studies of the Pb/Ag layers. Note that the underlying Ag layer is precisely the same Ag layer used in measurements Fig. 1 (b, e) There are three key observations: (I) Observation of "Ag-like" QWS despite that the Ag layer is buried underneath the Pb layer which is thicker than the escape depth (first reported by Brinkley *et al*.)[8]; although within the same energy window, the number of such QWS are roughly doubled; (II) the total absence of "Ag-like" QWS beyond a certain value of $k_\parallel$ (~ 0.6Å$^{-1}$); (III) observation of the folded back QWS that are the same as Pb/Si with the same Pb thickness.

The observation of "Ag-like" QWS near the zone center was first interpreted as coherent coupling of the QWS in the Ag underlayer to the Pb overlayer, thus enabling them to be observed despite being buried underneath in a depth more than the photoelectron escape depth (Fig. 1 (c, f)). We agree with the notion that the electronic states in Ag are coherently coupled to the Pb layers. Nevertheless, we interpret that such a coupling creates a new set of QWS for the Pb-Ag bimetallic layer as a whole and such QWS extend throughout the whole bi-metallic system. Thus, they can be detected by ARPES because the wavefunction is extended to the surface (Fig. 4a). In addition, the fact that the number of QWS within the same energy window is roughly doubled is also consistent with the interpretation that these are new QWS which are now confined within the whole bi-metallic thickness.

On the other hand, observations (II) and (III) indicate that in a certain region of $E - k$ space, the electronic states in the Ag and Pb layers stay within their respective regions without coupling. For example, for Ag QWS with large $k_\parallel$ (in region with $k_\parallel > 0.6$Å$^{-1}$) which can be clearly observed in the Ag/Si, are no longer detectable in the Pb/Ag system. Moreover, the folded back QWS in the Pb/Si system are now completely replicated in the Pb/Ag system.

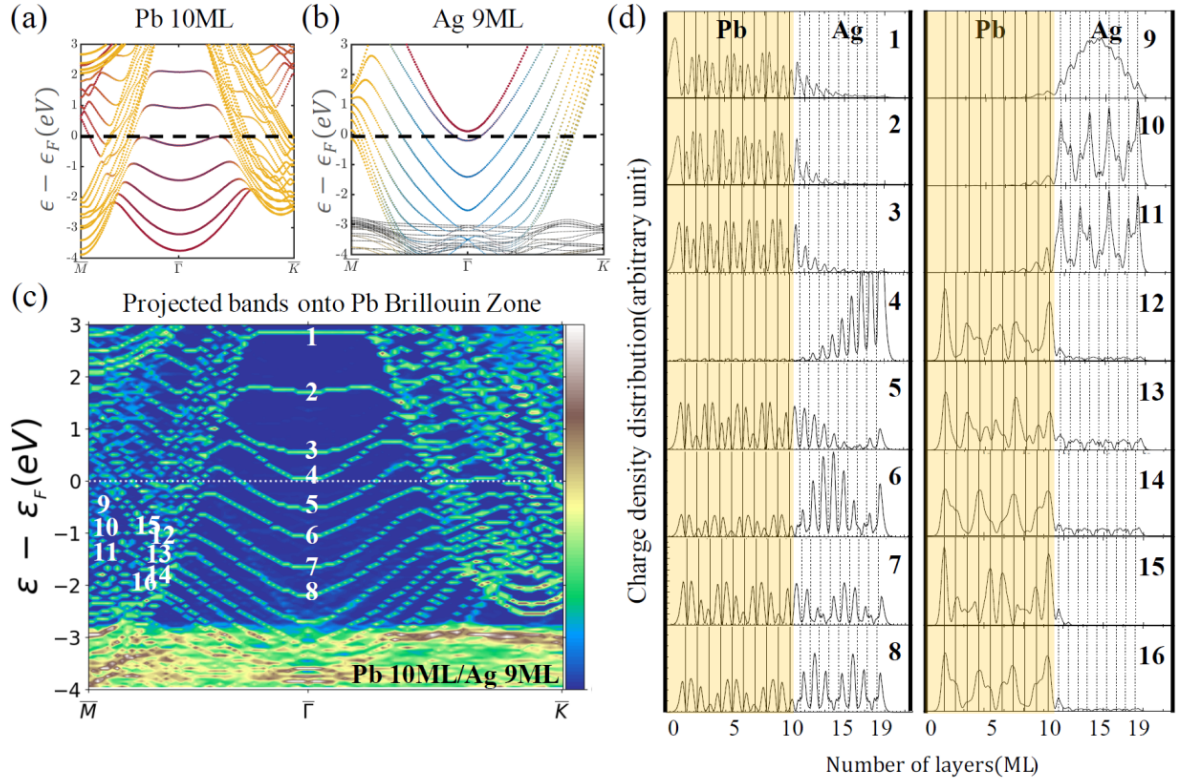

Figure 3. DFT calculational results of freestanding thin films (a) Pb 10ML, (b) Ag 9ML, and (c) Pb 10ML/ Ag 9ML. Blue, red and yellow color in (a) and (b) stands for contributions from $s, p_z$ and $p_x + p_y$ orbitals, respectively. Integrated charge-density distributions for states 1 to 16 marked in (c) are represented in (d). Solid and dashed vertical lines stand for Pb and Ag atomic planes, respectively.

We have further carried out DFT calculations to investigate this contrasting behavior. Namely in some regions of $E - k$ space, the electronic states in Pb and Ag layers are strongly coupled, forming a new set of QWS for the bi-metallic layer (observation (I)); whereas in other region of $E - k$ space, the electronic states in Pb and Ag layers are totally decoupled(observation II and III). Shown in Fig. 3(a ,b c) are DFT band structure calculations for 10 ML Pb, 9 ML Ag, and the 10/9 combination of the bimetallic Pb/Ag layers. To obtain electronic structures of the bimetallic layer (Fig. 3c), the lattice constant ratio between Pb and Ag is approximated by 6/5 which is within 1% of the actual ratio. The calculated electronic structures are then projected on to the Pb BZ for comparison. Fig. 3d shows the integrated electron density over atomic planes layer-by-layer for selected states marked on the band structures in Fig. 3c. State 1-3 are the empty states and can be projected only onto the Pb layer because they fall into the region of the L-gap in the Ag band structure. State 4 is the Ag surface state which exists there because the calculation is based on a free standing bi-metallic system with vacuum on the Ag side. Corresponding to our observation (I) are states 5-8 whose wavefunctions extend throughout the Ag and Pb layers. This calculation confirms our interpretation that for small $k_\parallel$ a new set of QWS are formed, which are confined within the Pb/Ag bi-metallic system as a whole.

For states 9-11, the wavefunctions are projected only onto the Ag layer. This is consistent with our observation (II) that Ag QWS with large $k_\parallel$ are retained only in the Ag layer without any coupling to the Pb layer. Meanwhile, for states 12-16, they are projected primarily onto the Pb regions albeit states 12-14 contain very small amplitudes in the Ag layer, suggesting a very weak coupling to the Ag states.

The experimental observations (I), (II) and (III), in conjunction with the DFT calculations firmly establish two regimes in the E-k space with contrasting behaviors for the Pb/Ag bi-metallic systems. In regions with a small $k_\parallel$ value the electronic states in Pb and Ag layers are strongly coupled, forming a new set of QWS for the bi-metallic system. This leads to an increase of the number of QWS whose energy locations are not just the replica of the QWS in individual layers. On the other hand, in regions where the Ag electronic has a large $k_\parallel$ value, or electronic states in Pb layer that are folded back (effectively a large $k_\parallel$ if viewed in the extended zone) then the states in Ag and Pb remain in their respective regions without coupling.

This interesting behavior can be understood by considering a simple model describing how electrons traverse across a potential step (Fig. 4a). For that, we use 5.5eV and 9.5eV as Ag and Pb fermi energy which are derived from free electron approximation (i.e. $E_F = \frac{\hbar^2}{2m}(3\pi^2 n)^{2/3}$ ). Though the fermi energy derived from DFT calculation and free electron approximation has small difference, it doesn't change the result. We can construct a model with a potential step $V_0$ of about 4 eV (with region 1 for Pb and region 2 for Ag). For simplicity, we consider a free electron traversing across this potential step (different effective mass calculation also shows similar behavior). The kinetic energy in region 1 is $E_1 = \frac{\hbar^2 k_1^2}{2m_0}$, and in region 2, $E_2 = \frac{\hbar^2 k_2^2}{2m_0}$, with $E_1 - E_2 = V_0$. When the electron is moving normal to the interface (i.e. $k_\parallel = 0$) and with enough energy to overcome the barrier, the transmission probability across the interface is simply $T = \frac{4k_1 k_2}{(k_1+k_2)^2}$.

Note that this transmission probability is rather large in most of the energy range. For example, at $E_1 - V_0 = 1\ eV$, $T = 0.82$ and quickly goes to 0.91 when $E_1 - V_0 = 2 eV$. In essence, the electronic interface is very transparent for small $k_\parallel$. In this case, electrons are confined within an effective thickness equivalent to the sum of the Pb and Ag thickness leading to a new set of QWS for the whole bi-metallic system. On the other hand, with a finite parallel momentum only the perpendicular component of the kinetic energy can be used to overcome the barrier, with $E_{1,\perp} = E_1 - \frac{\hbar^2 k_\parallel^2}{2m_0}$, and $E_{2,\perp} = E_2 - \frac{\hbar^2 k_\parallel^2}{2m_0}$, but the energy barrier remains the same. In this case, when $k_\parallel$ in Pb layer exceeds a critical value of $k_{\parallel,c} = \sqrt{\frac{2m_0(E_1-V_0)}{\hbar^2}}$ then the transmission goes to zero and total internal reflection occurs. For a typical value of $(E_1 - V_0) \approx 4 - 5.5\ eV$ and a free electron mass, this critical value would be roughly $1.0 - 1.2\ \text{Å}^{-1}$, which qualitatively agrees with our observation. Note also for the folded back bands with negative dispersion in Pb, they exist in the k-space between 0.5 and 0.9 Å$^{-1}$ which correspond to a $k_\parallel$ value of 1.17 to 1.57 Å$^{-1}$ in the extended zone, far exceeding the critical value for total internal reflection. This simple analysis explains why in the zone center, the states in Pb and Ag are strongly coupled together leading to integral electronic states for the bi-metallic layer as a whole. Whereas the folded back QWS in Pb layer are retained in the Pb regime and cannot couple to the electronic states in the Ag region due to total internal reflection. In the case of Ag QWS with large $k_\parallel$ (it has small $E_{2,\perp}$) as shown in Fig. 4c, the transmission drops very quickly (Fig. 4b) and the Pb and Ag electrons hardly interact with each other due to small transmission. Note that this is also the region where the Pb electronic structures show a big L gap, as illustrated in Fig 4a. Thus, either using a simple free electron transmission or a full band structure effect, it shows that Ag states in the large $k_\parallel$ region stay within the Ag layer underneath Pb and could not be detected.

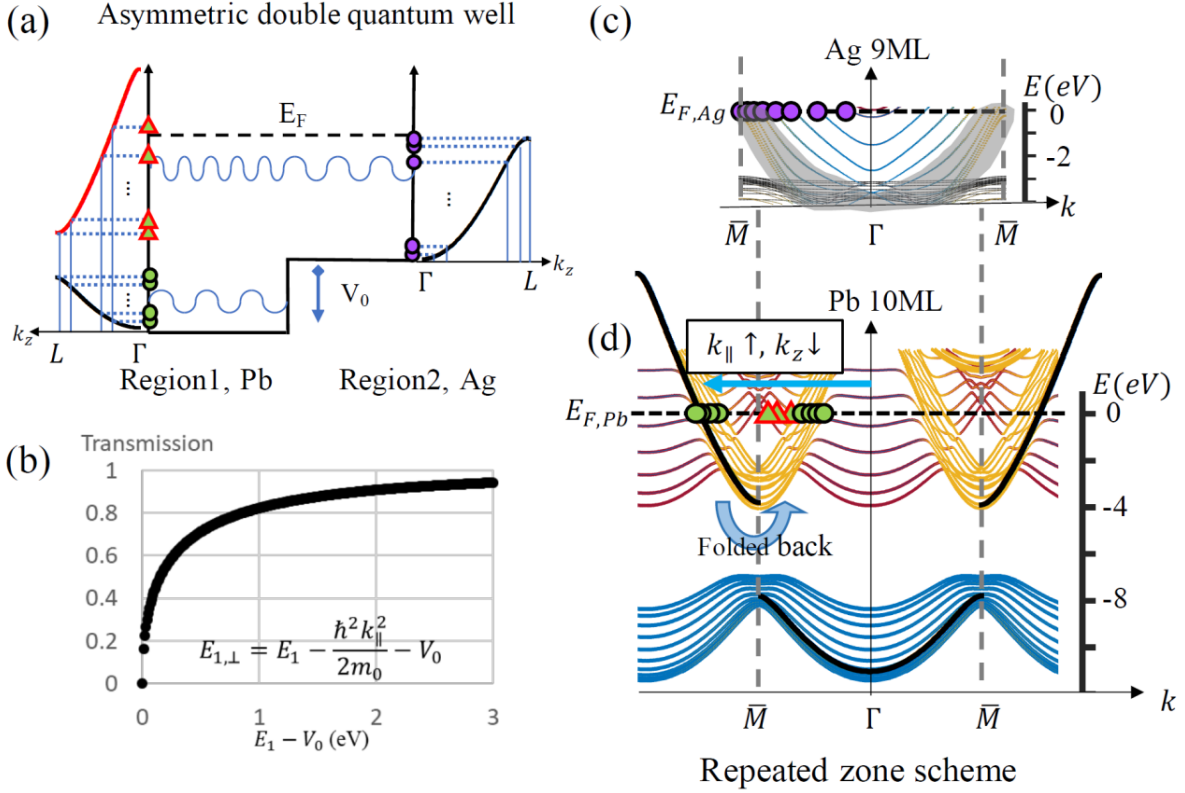

*Figure 4. Critical role of parallel momentum. (a) Schematic diagram of asymmetric double quantum well system, which consists of region1(Pb) and region2(Ag) with potential step $V_0$ on region2. Left- and Right-hand side E-k dispersions are drawn to show energy momentum quantization along 111 direction of individual Pb and Ag quantum wells. Green and purple markers represent Pb and Ag energy states. Red color marker outline stands for that the state comes from a folded back band. (b) The transmission probability at the Pb/Ag interface increases quickly after $E_1 > V_0$. (c) Ag band structure with markers. The markers indicate that each quantized band has different $k_z$ as shown in (a). Note that the bands in grey shaded area falls into Pb's L gap region, so they don't interact with Pb states. (d) Pb band structure is drawn in repeated zone scheme. Folded back states (green color marked) have large $k_\parallel$ in extended zone scheme (black solid lines) and small $k_z$. Due to this small $k_z$, it is not able to penetrate the potential step as shown in (a). On the other hand, the states with red outline marker which have large $k_z$ and relatively small $k_\parallel$ can penetrate the interface. So that, the red color parabolic shape bands can interact with Ag states in the composite QW system.*

These distinctly two different regions in k-space that influence the coupling between the QWS in Pb and Ag layers can also be viewed clearly in the constant energy surface (CES) mapping in k-space. Shown in Fig. 1 (b, c, d) are 2D k-space CES mapped near the Fermi energy for 9ML-Ag, 10ML-Pb/9ML-Ag, and 10ML-Pb on Si(111). For Ag, the CES near the Fermi energy appear as concentric rings extending to a maximum radius of $k_{max} \sim 1.25$ Å$^{-1}$. For Pb, near the Fermi energy the dominating CES are due to the folded back bands, which now appear as concentric hexagons. In the case of Pb/Ag/Si, in addition to these hexagons, two center rings are observed resulting from the composite quantum well state due to the coupling of Ag and Pb electronic states. Outside the hexagon rings, however, no trace of Ag-like states can be found, indicating that these states stay in the buried Ag layer deeper than the photoelectron escape depth and cannot be detected.

In summary, we have used ARPES to investigate quantum well state couplings in Pb/Ag systems. For the Pb/Ag bimetallic system, we discovered that there exist two distinct regimes in the $E\ vs.\ k$ phase space each with totally different coupling behavior. Near the zone center, within ~ 4 eV from the Fermi energy the electronic states of Pb and Ag are well coupled and form a new set of the QWS for the bimetallic system as a whole. On the other hand, for Ag QWS with large $k_\parallel$ the electronic states are retained within Ag layer only and are no longer detectable by ARPES. Similarly, the folded back QWS in the Pb/Si system are now completely replicated in the Pb/Ag system, indicating that these states are confined only in the Pb layer. When viewed in the extended zone, these folded back states correspond to states with very large $k_\parallel$, and experience total internal reflection. Our studies illustrate the important role played by the parallel momentum in determining the coupling of the electronic states in multi-layer quantum confined systems.

**Acknowledgement**
We acknowledge the funding from the Welch Foundation (F-1672), the Airforce (FA2386-18-1-4097), the NSF (DMR-1808751), the NSF MRSEC program (DMR-1720595), the NSF-EFMA-1542747 and the Academia Sinica Thematic Project (AS-TP-106-A07).

## Data availability

The data that support the findings of this study are available from the corresponding author upon reasonable request.

**Supplementary**

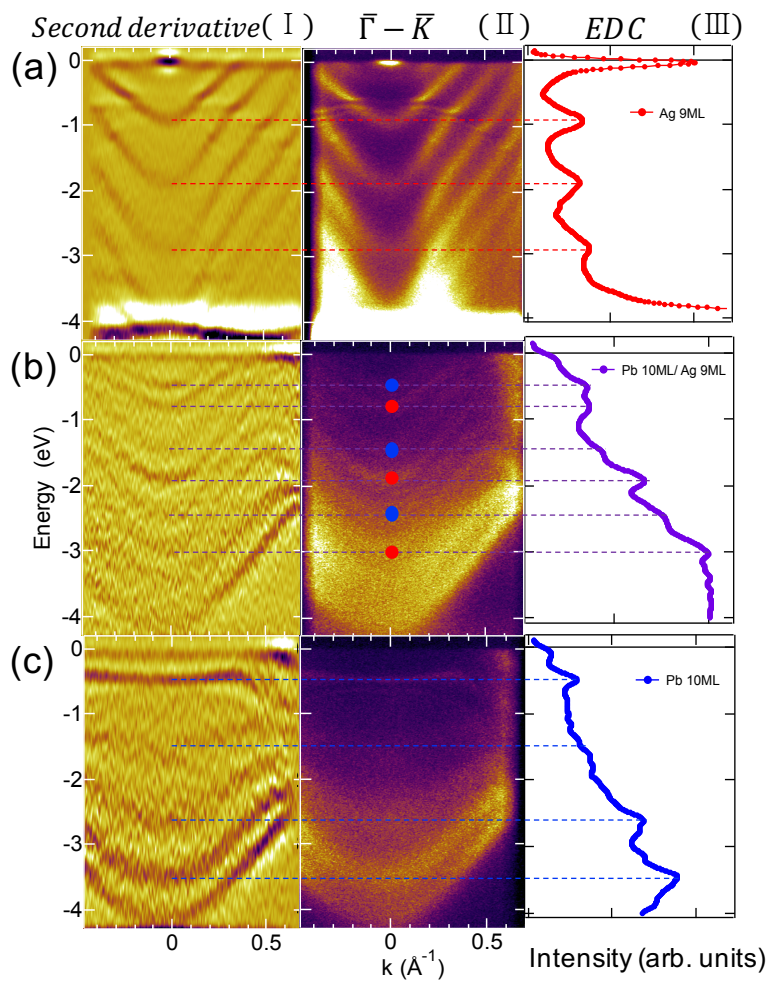

Figure S1. ARPES measurements on Ag(9ML), Pb(10ML)/Ag(9ML), Pb(10ML) using 40.8eV photon energy (He II). He II data shows better sensitivity on certain bands than He I 21.2eV photon energy.